\documentclass[12pt]{article}
\usepackage{amsfonts}
\usepackage{graphicx}
\newcommand{\adj}{{\bf adj}}
\newcommand{\rank}{{\bf rank}}

\newtheorem{theorem}{Theorem}
\newtheorem{corollary}{Corollary}

\def\qed{{\bf \hfill $\Box$}\endtrivlist}

\begin{document}

\begin{center}{\Large Metric Problems for Quadrics in Multidimensional Space} \end{center}

\begin{center}{\large Alexei Yu. Uteshev\footnote{alexeiuteshev@gmail.com} and Marina V. Yashina\footnote{marina.yashina@gmail.com}} \end{center}

\begin{center}{\it St. Petersburg State University, St. Petersburg, Russia} \end{center}

\begin{abstract}
Given the equations of the first and the second order surfaces in $\mathbb{R}^n$, our goal is to construct a univariate polynomial one of the zeros of which coincides with the square of the distance between these surfaces. To achieve this goal we employ Elimination Theory methods. The proposed approach is also extended for the case of parameter dependent surfaces.
\end{abstract}

\section{Introduction}
\label{Intro}
We solve the problem of finding the distance $d$ from the ellipsoid
\begin{equation}
X^T{\bf A}_1X+2B^T_1X-1=0
\label{eq1}
\end{equation}
either to linear surface given by the system of equations
\begin{equation}
C^T_1X=0,\dots,C^T_{k}X=0
\label{eq2}
\end{equation}
or to quadric
\begin{equation}
X^T{\bf A}_2X+2B^T_2X-1=0.
\label{eq3}
\end{equation}
Here $X=[x_1,\dots,x_n]^T$ is the column of variables, $\{B_1,B_2,C_1,\dots,C_{k}\} \subset \mathbb R^n$ are the given columns, ${\bf A}_1$ and ${\bf A}_2$ are the given symmetric matrices and ${\bf A}_1$ is sign-definite.

The distance is evaluated in Euclidean metrics $|| \cdot ||_2$, i.e. $$d= \min \sqrt{(X-Y)^T(X-Y)}$$
subject to $\{ X \in \mathbb R^n | X^T{\bf A}_1X+2B_1^TX-1=0 \}$ and $\{ Y \in \mathbb R^n | C_1^TY=0, \dots, C_k^TY=0 \}$ or $\{ Y \in \mathbb R^n | Y^T{\bf A}_2Y+2B_2^TY-1=0 \}$.

Being a problem of nonlinear optimization it can be solved via generation of a suitable iterative procedures \cite{b2},\cite{b1} or by application of some symbolic transformation of algebraic equations aiming at reducing the number of involved variables.
Thus, for instance, for the distance problem between (\ref{eq1}) and (\ref{eq3}) the starting point is the following system resulted from the Lagrange multipliers method
\begin{equation}
\left\{
\begin{array}{l}
X-Y-\lambda_1({\bf A}_1X+B_1)={\mathbb O}, \ -X+Y-\lambda_2({\bf A}_2Y+B_2)={\mathbb O}\\
X^T{\bf A}_1X+2B_1^TX=1, \ Y^T{\bf A}_2Y+2B_2^TY=1.
\end{array}
\right.
\label{eq4}
\end{equation}

Several publications \cite{b4}, \cite{b2} were devoted to the development of mentioned approach\-es for the dimensions $n=2$ and $n=3$. They were focused on finding the coordinates $X$ and $Y$ of the nearest points in the considered surfaces. In comparison with those approaches, in the present paper we suggest an alternative one aimed first at evaluation of the distance itself. This is achieved via introduction of a new variable $z$ by the equation
$$z-(X-Y)^T(X-Y)=0.$$
Being attached to the system (\ref{eq4}) this equation provides the critical values of the distance function. For the obtained algebraic system one may apply an algebraic procedure of elimination of all the variables except for $z$. That means, it is possible to construct an algebraic univariate equation ${\cal F}(z)=0$ one of the zeros of which (generically minimal positive) coincides with the square of the distance we are looking for. This construction can be performed either via the Gr\"{o}bner basis computation or with the aid of the classical Elimination Theory toolkit. We have chosen the second approach and succeeded in finding explicit expressions for the polynomial ${\cal F}$ for each of the stated problems in $n-$dimensional space. Any of the real zeros of ${\cal F}(z)=0$ corresponds to a pair of points on the treated surfaces, and we also suggest an algorithm for evaluation of their coordinates. It turns out that these coordinates can be generically expressed as rational functions of the value of $z$. We also treat a surface intersection problem. Some of results from Sections \ref{DisQL} and \ref{DisQQ} were first formulated in \cite{b10}. In the present paper we give a proof for Theorem \ref{t2} (missed in \cite{b10}) and correct one void in the proof of Theorem \ref{t1}.

\section{Algebraic preliminaries}
\label{AlgPr}
From the mentioned in the previous paragraph Elimination Theory toolkit, the most perfect gadget for our purpose turns out to be the discriminant. We will be in need of its univariate and bivariate form.

{\bf Univariate discriminant.}
For the univariate polynomial $g(x)=b_0 x^N + b_1 x^{N-1} + \dots + b_N \in \mathbb{C}[x]$, $b_0 \ne 0$, $N \ge 2$ its discriminant is formally defined as
\begin{equation}
{\cal D}_x(g) \stackrel{def}{=} b_0^{N-1} \prod_{j=1}^N g'(\mu_j), \label{2_1}
\end{equation}
where $\{\mu_1, \dots, \mu_N \}$ is a set of zeros of $g(x)$ counted in accordance with their multiplici\-ties. We will also use an alternative definition of discriminant
\begin{equation}
{\cal D}_x(g) \stackrel{def}{=} (-1)^{N(N-1)/2} N^N b_0^{N-1} \prod_{j=1}^{N-1} g(\lambda_j), \label{disk1}
\end{equation}
where $\{ \lambda_1, \dots, \lambda_{N-1}\}$ is a set of zeros of $g'(x)$ counted in accordance with their multiplicities. The constructive computation of discriminant -- in the form of polyno\-mial function of the coefficients of $g(x)$ -- can be performed with the aid of several determinantal representations. We will utilize the B\'{e}zout's approach \cite{b5} which is based on the coefficients of the remainders on dividing $x^\ell g(x)$ by $g'(x)$:
$$x^\ell g(x) \equiv b_{\ell 0}+b_{\ell 1}x+ \dots + b_{\ell,N-2}x^{N-2} + q_\ell(x) g'(x), \ q_\ell(x) \in \mathbb{C}[x]$$
for $\ell \in \{ 0, \dots, N-2 \}.$ Compose the matrix from these coefficients
\begin{equation}
{\mathfrak B} \stackrel{def}{=}[b_{\ell j}]_{\ell,j=0}^{N-2}. \label{BB}
\end{equation}
Denote by ${\mathfrak B}_{N-1,j}$ the cofactor to the corresponding entry of the last row of ${\mathfrak B}$.

\begin{theorem}
One has $${\cal D}_x(g)=N^N b_0^{N-1} \det {\mathfrak B}.$$
The polynomial $g(x)$ possesses a multiple zero iff $\det {\mathfrak B} =0$. Under this condition, the multiple zero is unique iff ${\mathfrak B}_{N-1,1} \ne 0$; in this case it can be expressed rationally via the coefficients of $g(x)$:
\begin{equation}
\lambda = \frac{{\mathfrak B}_{N-1,2}}{{\mathfrak B}_{N-1,1}}. \label{lam}
\end{equation}
\label{t21}
\end{theorem}
{\it Example.} Find the real values of the parameter $\alpha$ under which the polynomial
$$g(x)=x^5+6\,x^4+2\,x^3+\alpha \,x^2-x+3$$
possesses a multiple zero, and evaluate this zero.

\noindent {\it Solution.} We compute first the remainders on division of $g,\, xg,\, x^2g,\, x^3g$ by $g'(x)$:
\begin{eqnarray*}
&& \frac{81}{25}  + \left( - \frac{12\,\alpha }{25}- \frac{4}{5} \right) \,x + \left( \frac {3\,\alpha }{5} - \frac{36}{25} \right) \,x^{2} - \frac {124}{25} \, \,x^{3}, \\
&&\\
&-& \frac {124}{125} + \left( \frac {248\, \alpha }{125} + \frac {81}{25} \right) \,x + \left( -\frac {12\,\alpha}{25}  + \frac {644}{125} \right) \,x^{2} + \left( \frac {3\,\alpha }{5} + \frac {2796}{125} \right) \,x ^{3}, \\
&&\\
&& \frac {3\,\alpha }{25}+ \frac {2796}{625}+  \left( - \frac {6}{25} \,\alpha ^{2} - \frac {5592}{625} \,\alpha - \frac {124}{125} \right)\,x + \left(\frac {158\, \alpha }{125} - \frac {14751}{625} \right)\,x^{2}  \\
&+&\left( - \frac{84\,\alpha }{25}- \frac {63884}{625} \right)\,x^{3}, \\
&&\\
&-& \frac {84\,\alpha }{125} -\frac {63884}{3125}+ \left( \frac {168}{125} \,\alpha ^{2} + \frac{128143}{3125} \,\alpha + \frac {2796}{625} \right) \,x \\
&+&\left(- \frac {6}{25} \,\alpha^{2} - \frac {3072}{625} \,\alpha  + \frac {380204}{3125} \right)\,x^{2} + \left( \frac {2174\,\alpha }{125}+ \frac {1459461}{3125} \right)\,x^{3}.
\end{eqnarray*}
Then compose the matrix $\mathfrak B$ from the coefficients of powers of $x$ and compute its determinant
$$\det \mathfrak B = \frac {(\alpha  + 7)\,(324\,\alpha ^{4} + 5481\, \alpha ^{3} - 87771\,\alpha ^{2} - 409817\,\alpha  + 5759315)}{3125}.$$
The discriminant ${\cal D}_x(g)$ coincides (up to a numerical factor) with the numerator of the last fraction and it vanishes iff
$$\alpha \in \{-24.63939477, \,-9.29644677, \,-7\}.$$
To evaluate the corresponding multiple zero of $g(x)$, we utilize formula (\ref{lam}):
$$\lambda =  - {\displaystyle \frac { {\displaystyle \frac {27}{625} \,\alpha ^{3} + \frac {18}{5} \, \alpha ^{2} + \frac {32537}{625}\,\alpha  + \frac {2724}{625}}}{{\displaystyle - \frac {54}{625} \,\alpha^{4} - \frac {1296}{625}\, \alpha ^{3} + \frac {4508}{625}\,\alpha ^{2} + \frac {17208}{125}\,\alpha  - \frac {57532}{625}}}} \ ,$$
where numerator and denominator are just the minors to the entries of the last row of $\mathfrak B$. Substitution of the obtained values of $\alpha$ into this formula yields the corresponding values of multiple zeros: $-3.80947138, \ 0.74648466, \ -1$.

\begin{corollary} Let $\phi(x)=p(x)/q(x)$ be rational function with relatively prime $p(x)$ and $q(x)$. Functions $\phi(x)$ and $\phi'(x)$ posses a common zero iff ${\cal D}_x(p)=0$.
\label{colfrac}
\end{corollary}
{\bf Proof.} One has $\phi(x)=0$ iff $p(x)=0$. Let $\deg p(x) =m$ and $\lambda_1, \dots, \lambda_m$ stand for the zeros of $p(x)$. Thus
$$\phi'(x)=\frac{p'(x)}{q(x)}-\frac{p(x)q'(x)}{q^2(x)} \Rightarrow \phi'(\lambda_j) = \frac{p'(\lambda_j)}{q(\lambda_j)} \mbox{ for } j \in \{1,\dots,m\};$$
here $q(\lambda_j) \ne 0$ under the assumption of the corollary. Therefore
$$\prod_{j=1}^m \phi'(\lambda_j) = \prod_{j=1}^m p'(\lambda_j) \Big / \prod_{j=1}^m q(\lambda_j)$$
and in accordance with the definition (\ref{disk1}) this product vanishes iff ${\cal D}_x(p)=0$. \qed

\begin{corollary} For polynomial $g(x)$ of degree $N \ge 2$ and a constant $A \in \mathbb{C}$ one has:
\begin{eqnarray}
{\cal D}_x(A \cdot g(x))&=& A^{2N-2} {\cal D}_x(g), \label{unidiskcor1} \\
{\cal D}_x(x \cdot g(x)) &=& [g(0)]^2 {\cal D}_x(g). \label{unidiskcor2}
\end{eqnarray}
\end{corollary}
\begin{theorem}
One can find polynomials providing the so-called linear repre\-sentation of the discriminant, i.e., the pair $\{ u(x), \, v(x) \} \subset \mathbb C [x]$ satisfying the identity
\begin{equation}
v(x)g(x)+u(x)g'(x) \equiv \det \mathfrak B. \label{lindisk}
\end{equation}
Here $v(x)$ can be represented as the determinant of the matrix obtained on replacing the first column of $\mathfrak B$ by $[1,x,\dots,x^{N-2}]^T$, while
$$u(x)=-\displaystyle{\frac{1}{N} \left( x+ \frac{1}{N} \frac{b_1}{b_0} \right) v(x)- \frac{1}{N b_0} \det \widehat {\mathfrak B}},$$
where $\widehat {\mathfrak B}$ denote the matrix obtained from $\mathfrak B$ by replacing its first column by
\begin{eqnarray*}
&&[0,\, b_{0,N-2},\, b_{0,N-2}x+b_{1,N-2},\, b_{0,N-2}x^2+b_{1,N-2}x+b_{2,N-2}, \dots, \\ &&b_{0,N-2}x^{N-3}+b_{1,N-2}x^{N-4}+ \dots+b_{N-3,N-2}]^T.
\end{eqnarray*}
The polynomials $u(x)$ and $v(x)$ satisfy the restrictions
$$\deg u < N-1, \ \deg v < N-2.$$ \label{teor24}
\end{theorem}

{\bf Bivariate discriminant.}
For the given polynomial $g(X) \in \mathbb C[X], \ X=(x_1,x_2),$ $\deg g =N \ge 2$ we define its discriminant as
$${\cal D}_X(g) \stackrel{def}{=} \prod_{j=1}^{\mathfrak N} g(\Lambda_j).$$
Here $\Lambda_j = (\lambda_{j_1}, \lambda_{j_2}) \in \mathbb C^2$ stands for the stationary point of $g(X)$, i.e. a zero of the system $\partial g/ \partial x_1 =0, \ \partial g/ \partial x_2=0.$ In generic case, the latter possesses precisely $\mathfrak N =(N-1)^2$ (B\'{e}zout's number) zeros in $\mathbb{C}^2$. Constructive computation of ${\cal D}_X(g)$ is possible with the aid of an analogue to the division process utilizied in the univariate case. Choose the set of $\mathfrak N$ power products in $X$:
\begin{equation}
\left\{ {\cal M}_\ell(X) \right\}_{\ell=0}^{\mathfrak N -1} = \left\{ x_1^{j_1}x_2^{j_2} \big| \, 0 \le j_1 < N -1,
 0 \le j_2 \le 2(N-j_1-2) \right\}. \label{eq6}
\end{equation}
For instance, one has for $N=5$:
\begin{equation}
\left \{ {\cal M}_\ell(X) \right \}_{\ell=0}^{15} =
\begin{array}{lrrrrll}
\{ 1, & x_2, & x_2^2, & x_2^3, & x_2^4, & x_2^5, & x_2^6, \\
\phantom{\{} x_1, & x_1x_2, & x_1x_2^2, & x_1x_2^3, & x_1x_2^4, & &\\
\phantom{\{} x_1^2, & x_1^2x_2, & x_1^2x_2^2, & & & &\\
\phantom{\{} x_1^3 \, \}. & & & & & & \end{array} \label{stepeni}
\end{equation}
We will call the reduction of the polynomial ${\cal M}_\ell(X)g(X)$ modulo $\partial g/ \partial x_1$ and $\partial g/ \partial x_2$ its representation in the form
\begin {eqnarray}
{\cal M}_\ell(X)g(X) &\equiv& b_{\ell 0}{\cal M}_0(X) + \dots + b_{\ell, \mathfrak N-1}{\cal M}_{\mathfrak N-1}(X)   \label{eq61}\\
&&  + \,q_{\ell 1}(X) \partial g/\partial x_1 + q_{\ell 2}(X) \partial g/ \partial x_2, \nonumber
\end{eqnarray}
with $\{q_{\ell 1}(X),q_{\ell 2}(X)\} \subset \mathbb{C}[X].$
Theoretical possibility of such a representation as well as constructive algorithms for its implementation are discussed in \cite{b5}. We note just only that in case of reducibility, the coefficients $b_{\ell j}$ can be expressed as rational functions of the coefficients of $g(X)$. Reorder the set (\ref{eq6}) in such a manner that ${\cal M}_0=1, {\cal M}_1=x_1, {\cal M}_2=x_2$ and make the matrix from the coefficients of the reductions (\ref{eq61}) for $\ell \in \{ 0, \dots, \mathfrak N -1\}$, i.e. for all the power products from (\ref{eq6}):
\begin{equation}
{\mathfrak B} = \left[ b_{\ell j} \right]_{\ell, j=0}^{\mathfrak N -1}. \label{BB2}
\end{equation}
Denote by ${\mathfrak B}_{\mathfrak N,j}$ the cofactor to the corresponding entry of the last row of ${\mathfrak B}$.
\begin{theorem} One has
$${\cal D}_X(g)=\det {\mathfrak B}.$$
The polynomial $g(X)$ possesses a multiple zero $\Lambda=(\lambda_1, \lambda_2) \in \mathbb C^2$ (i.e. the zero for which $g=0, \, \partial g/ \partial x_1=0, \, \partial g/ \partial x_2=0$) iff $\det {\mathfrak B}=0$. Under this condition, the multiple zero is unique if ${\mathfrak B}_{\mathfrak N,1} \ne 0$; in this case it can be expressed as
\begin{equation}
\lambda_1={\mathfrak B}_{\mathfrak N,\,2}/{\mathfrak B}_{\mathfrak N,\,1}, \ \lambda_2={\mathfrak B}_{\mathfrak N,\,3}/{\mathfrak B}_{\mathfrak N , \, 1}. \label{eq7}
\end{equation} \label{tva}
\end{theorem}
{\bf Schur formula.} Subsequently we will frequently use the following Schur complement formula for the determinant of a block matrix \cite{b6}:
\begin{equation}
\det \left(
\begin{array}{cc}
{\bf U} &  {\bf V} \\
{\bf S} & {\bf T}
\end{array}
\right) = \det {\bf U} \det \left( {\bf T} - {\bf S} {\bf U}^{-1}{\bf V}
\right),
\label{Schur}
\end{equation}
here ${\bf U}$ and ${\bf T}$ are square matrices and ${\bf U}$ is non-singular.

\section{Distance between a quadric and a linear surface}
\label{DisQL}

We treat the equa\-tions of the surfaces in the form (\ref{eq1}) and (\ref{eq2}) and assume the columns $C_1, \dots, C_k$ to be linearly independent (the latter results in the restriction $k \le n$). Compose the matrices ${\bf C} \stackrel {def}{=} [C_1, \dots, C_k]$ and
\begin{equation}
{\bf G} \stackrel{def}{=} {\bf C}^T {\bf C}, \label{grammatr}
\end{equation}
i.e. ${\bf G}$ is the Gram matrix for the columns $C_1, \dots, C_k$. Due to imposed restriction on $C_1, \dots, C_k$, the matrix ${\bf G}$ is nonsingular.
\begin{theorem} The condition
\begin{equation}
0 \le \left|
\begin{array}{ccc}
{\bf A}_1 & B_1 & {\bf C}\\
B_1^T & -1 & {\mathbb O}\\
{\bf C}^T & {\mathbb O} & {\mathbb O}
\end{array} \right| \times
\left\{ \begin{array}{l}
(-1)^{k-1}, \ if \ {\bf A}_1 \mbox{ is positive definite}, \\
(-1)^n, \ if \ {\bf A}_1 \mbox{ is negative definite}
\end{array} \right.
\label{inter}
\end{equation}
is the necessary and sufficient one for the linear surface (\ref{eq2}) to intersect the ellipsoid (\ref{eq1}); in this case one has $d=0$. If this intersection condition is not satisfied then the value $d^2$ coincides with the minimal positive zero of the equation
\begin{equation}
{\cal F}(z) \stackrel{def}{=}{\cal D}_\mu \left( \mu^k \left|
\begin{array}{ccc}
{\bf A}_1 & B_1 & {\bf C}\\
B_1^T & -1 + \mu z & {\mathbb O}\\
{\bf C}^T & {\mathbb O} & \displaystyle{\frac{1}{\mu}} {\bf G}
\end{array}
\right| \right)=0
\label{eq8}
\end{equation}
provided that this zero is not a multiple one.
\label{t1}
\end{theorem}

{\bf Proof.} {\bf I. Finding the intersection condition.}
Let us first find the critical value of\footnote{To simplify the notation we will type matrices ${\bf A}$ and $B$ without their indices.}
$V(X)=X^T{\bf A}X+2B^TX-1$ in the surface ${\bf C}^TX={\mathbb O}$.
The critical point of the Lagrange function
$$X^T{\bf A}X+2B^TX-1-\nu_1 C^T_1X-\dots-\nu_k C^T_kX$$
satisfies the system of equations
$$2{\bf A}X+2B-{\bf C}\left[ \nu_1, \dots,\nu_k \right]^T= {\mathbb O},\ {\bf C}^T X= {\mathbb O} \enspace .$$
Therefore
\begin{equation}
X=-{\bf A}^{-1}B+\frac{1}{2}{\bf A}^{-1}{\bf C}\left[ \nu_1, \dots,\nu_k \right]^T
\label{alpha}
\end{equation}
with
\begin{equation}
\left[ \nu_1, \dots,\nu_k \right]^T = 2 \left( {\bf C}^T{\bf A}^{-1}{\bf C} \right)^{-1} {\bf C}^T{\bf A}^{-1}B \enspace .
\label{beta}
\end{equation}
Substitution of (\ref{beta}) into  (\ref{alpha}) yields
$$X_e=-{\bf A}^{-1}B+{\bf A}^{-1}{\bf C}\left( {\bf C}^T{\bf A}^{-1}{\bf C} \right)^{-1}{\bf C}^T{\bf A}^{-1}B $$
and the corresponding critical value of $V(X)$ subject to ${\bf C}^T X= {\mathbb O}$ equals
$$ V(X_e)= -(B^T{\bf A}^{-1}B+1-B^T{\bf A}^{-1}{\bf C}({\bf C}^T{\bf A}^{-1}{\bf C})^{-1} {\bf C}^T{\bf A}^{-1}B) \enspace .$$
With the aid of Schur formula (\ref{Schur}) one can transform the last expression into
\begin{equation}
V(X_e)=
\frac{-\left |
\begin{array}{cc}
{\bf C}^T{\bf A}^{-1}{\bf C} & {\bf C}^T{\bf A}^{-1}B\\
B^T{\bf A}^{-1}{\bf C} & B^T{\bf A}^{-1}B+1
\end{array}
\right | }{\det({\bf C}^T{\bf A}^{-1}{\bf C})}=
\frac{ (-1)^k \left |
\begin{array}{ccc}
{\bf A} & B & {\bf C}\\
B^T & -1 & {\mathbb O}\\
{\bf C}^T & {\mathbb O} & {\mathbb O}
\end{array}
\right |}{\det({\bf A}) \det({\bf C}^T{\bf A}^{-1}{\bf C})} \enspace .
\label{gamma}
\end{equation}
If $V(X_e)=0$ then the linear surface (\ref{eq2}) is tangent to the ellipsoid (\ref{eq1}) at $X=X_e$. Otherwise let us compare the sign of $V(X_e)$ with the sign of $V(X)$ at infinity. These signs will be distinct iff the considered surfaces intersect. If ${\bf A}$ is positive definite then
$V_\infty>0$, $\det({\bf A})>0$ and $\det({\bf C}^T{\bf A}^{-1}{\bf C})>0$. Therefore, $V(X_e)<0$ iff the numerator in (\ref{gamma}) is negative. This confirmes (\ref{inter}). The case of negative definite matrix ${\bf A}$ is treated similarly.

{\bf II. Distance evaluation.}
Using the Lagrange multipliers method we reduce the constrained optimization problem to the following system of algebraic equations
\begin{eqnarray}
&&X-Y-\lambda {\bf A}X-\lambda B={\mathbb O} \label{w1}\\
&&X-Y+\frac{1}{2}{\bf C}[\lambda_1, \dots, \lambda_k]^T={\mathbb O} \label{w2}\\
&&X^T{\bf A}X+2B^TX-1=0 \label{w3}\\
&&{\bf C}^TY={\mathbb O} \enspace . \label{w4}
\end{eqnarray}
We introduce also a new variable responsible for the critical values of the distance function:
\begin{equation}
z-(X-Y)^T(X-Y)=0 \enspace . \label{w5}
\end{equation}
Our aim is to eliminate all the variables from the system (\ref{w1})--(\ref{w5}) except for $z$. We express first $X$ and $Y$ from (\ref{w1}) and (\ref{w2}) (hereinafter ${\bf I}$ stands for the identity matrix of an appropriate order):
\begin{eqnarray}
X&=&-{\bf A}^{-1}B-\displaystyle{\frac{1}{2\lambda}}{\bf A}^{-1}{\bf C}[\lambda_1, \dots, \lambda_k]^T \label{w6}\\
Y&=&-{\bf A}^{-1}B-\displaystyle{\frac{1}{2\lambda}}({\bf A}^{-1}-\lambda {\bf I}) {\bf C}[\lambda_1, \dots, \lambda_k]^T. \label{w7}
\end{eqnarray}
Then we substitute (\ref{w7}) into (\ref{w4}) with the aim to express $\lambda_1, \dots, \lambda_k$ via $\lambda$. This can be performed with the aid of the following matrix
\begin{equation}
{\bf M} \stackrel{def}{=}\displaystyle{\frac{1}{\lambda}}{\bf C}^T{\bf A}^{-1}{\bf C}-
{\bf C}^T{\bf C}= \mu{\bf C}^T{\bf A}^{-1}{\bf C}-{\bf G},
\label{kappa}
\end{equation}
with ${\bf G}$ defined by (\ref{grammatr}) and $\mu \stackrel{def}{=}1/\lambda$. Indeed, one has
\begin{equation}
{\bf M}[\lambda_1, \dots, \lambda_k]^T=-2 {\bf C}^T{\bf A}^{-1}B \label{w8}
\end{equation}
and, provided that ${\bf M}$ is non-singular,
\begin{equation}
[\lambda_1, \dots, \lambda_k]^T=-2{\bf M}^{-1}{\bf C}^T{\bf A}^{-1}B. \label{w9}
\end{equation}
Now substitute (\ref{w9}) into (\ref{w2}) and then the obtained result into (\ref{w5}):
\begin{equation}
z-B^T{\bf A}^{-1}{\bf C}{\bf M}^{-1}{\bf G}{\bf M}^{-1}{\bf C}^T{\bf A}^{-1}B=0. \label{w10}
\end{equation}
Equation (\ref{w10}) is a rational one with respect to the variables $\mu$ and $z$.

To find an extra equation for these variables, let us transform (\ref{w3}) using (\ref{w6}) and (\ref{w9})
\begin{eqnarray*}
0&=& X^T{\bf A}X+2B^TX-1  \\
&=& -B^T{\bf A}^{-1}B-1+ \mu B^T{\bf A}^{-1}{\bf C}{\bf M}^{-1}(\mu {\bf C}^T{\bf A}^{-1} {\bf C}-{\bf G}+{\bf G}){\bf M}^{-1}{\bf C}^T{\bf A}^{-1}B.
\end{eqnarray*}
Using (\ref{kappa}) and (\ref{w10}), the last equation takes the form
\begin{equation}
\Psi(\mu,z) \stackrel{def}{=} -1+\mu z-B^T{\bf A}^{-1}B+ \mu B^T{\bf A}^{-1}{\bf C}{\bf M}^{-1}{\bf C}^T{\bf A}^{-1}B=0. \label{w11}
\end{equation}
Therefore, system (\ref{w1})--(\ref{w5}) is reduced to (\ref{w10})--(\ref{w11}). It can be verified that the left-hand side of (\ref{w10}) is just the derivative with respect to $\mu$ of that of (\ref{w11}) and, thus, it remains to eliminate $\mu$ from the system
$$\Psi(\mu,z)=0, \ \Psi'_{\mu}(\mu,z)=0.$$
Taking into account Corollary \ref{colfrac} from Sect. \ref{AlgPr}, one can perform this with the aid of discriminant -- and that is the reason for its appearence in the statement of the theorem.

Schur formula (\ref{Schur}) helps once again in representing $\Psi(\mu, z)$ in the determinantal form:
\begin{equation}
\Psi(\mu,z) \equiv
\frac{\left |
\begin{array}{ccc}
{\bf A} & B& {\bf C}\\
B^T & -1+\mu z & {\mathbb O}\\
{\bf C}^T& {\mathbb O} & \displaystyle{\frac{1}{\mu}}{\bf G}
\end{array}
\right |}
{\left |
\begin{array}{cc}
{\bf A} & {\bf C}\\
{\bf C}^T & \displaystyle{\frac{1}{\mu}} {\bf G}
\end{array}
\right |}=
\frac{\mu^k \left |
\begin{array}{ccc}
{\bf A} & B& {\bf C}\\
B^T & -1+\mu z & {\mathbb O}\\
{\bf C}^T& {\mathbb O} & \displaystyle{\frac{1}{\mu}}{\bf G}
\end{array}
\right |}
{\det({\bf A}) \det({\bf M})} \enspace.
\label{w12}
\end{equation}

{\bf III. Finding the nearest points on the surfaces.}
Once the real zero $z=z_\ast$ of (\ref{eq8}) is evaluated, one can reverse the elimination scheme from part II of the proof in order to find the corresponding points $X_\ast$ and $Y_\ast$ on the surfaces.

For $z=z_\ast$, the polynomial in $\mu$ standing in the numerator of (\ref{w12})
\begin{equation}
\Phi(\mu,z) \stackrel{def}{=}\mu^k \left|
\begin{array}{ccc}
{\bf A} & B & {\bf C} \\
B^T & -1+\mu z & \mathbb{O} \\
{\bf C}^T & \mathbb{O} & \displaystyle{\frac{1}{\mu}} {\bf G}
\end{array} \right| \label{fi}
\end{equation}
has a multiple zero $\mu=\mu_\ast$. Provided that the multiple zero is unique, it can be expressed rationally in terms of the coefficients of this polynomial (and consequently in $z_\ast$) with the aid of (\ref{lam}). We substitute this value into (\ref{kappa}) then resolve the linear system (\ref{w8}) with respect to $\lambda_1, \dots, \lambda_k$ and, finally, substitute the obtained values into (\ref{w6}) and (\ref{w7}).

However, this algorithm fails if for $\mu=\mu_\ast$ the matrix ${\bf M}$ becomes singular. For explanation of the geometrical reason, one may recall that the distance between the surfaces may be attained not in a unique pair of points.

We avoid this case by imposing the simplicity restriction for the minimal zero of ${\cal F}(z)$ in the statement of the theorem.

{\bf IV. Nonsingularity of the matrix ${\bf M}$.}
In accordance with Theorem \ref{teor24}, the polyno\-mial ${\cal F}(z)$, being the discriminant of $\Phi(\mu,z)$, permits the linear representa\-tion
\begin{equation}
{\cal F}(z) \equiv v(\mu,z) \Phi +u(\mu,z) \Phi_\mu',
\label{krat1}
\end{equation}
with the polynomials $\{ v(\mu,z),u(\mu,z)\} \subset \mathbb{R}[\mu,z]$ satisfying the degree restrictions:
$\deg_\mu u< \deg_\mu \Phi, \deg_\mu v < \deg_\mu \Phi '_\mu.$

If $z=z_\ast$ stands for the zero of ${\cal F}(z)$, then $\Phi(\mu,z_\ast)$ and $\Phi_\mu'(\mu,z_\ast)$ possesses a common zero $\mu=\mu_\ast$. Differentiate (\ref{krat1}) with respect to $z$:
$$ {\cal F}'(z) \equiv v_z' \Phi + v \Phi_z' + u_z' \Phi_\mu' + u \Phi_{\mu z}''$$
and substitute $\mu=\mu_\ast$, $z=z_\ast$:
\begin{equation}
{\cal F}'(z_\ast) = v \Phi_z'+u \Phi_{\mu z}''.
\label{krat2}
\end{equation}
We intend to prove that $u(\mu_\ast,z_\ast)=0$. For this aim, differentiate (\ref{krat1}) with respect to $\mu$:
$$0 \equiv v_\mu' \Phi +v \Phi_\mu'+ u_\mu' \Phi_\mu' + u \Phi_{\mu^2}''$$
and substitute $\mu=\mu_\ast$, $z=z_\ast$
\begin{eqnarray}
&& \left. 0=u(\mu_\ast,z_\ast) \frac{\partial^2 \Phi}{\partial \mu^2} \right|_{(\mu_\ast,z_\ast)} \Leftrightarrow \label{22*} \\
&& u(\mu_\ast,z_\ast)=0 \qquad \mbox{ or } \qquad \left. \displaystyle{\frac{\partial^2 \Phi}{\partial \mu^2}} \right|_{(\mu_\ast,z_\ast)}=0. \label{krat3}
\end{eqnarray}
The second alternative from (\ref{krat3}) has the meaning that the zero $\mu=\mu_\ast$ is of multiplicity $k$ greater than 2 for $\Phi(\mu,z_\ast)$. In this case, one has from (\ref{krat1}):
$$0 \equiv v(\mu,z_\ast) \Phi(\mu,z_\ast) + u(\mu,z_\ast) \Phi_{\mu}'(\mu,z_\ast) \Leftrightarrow$$
\begin{equation}
u(\mu,z_\ast) \Phi_\mu'(\mu,z_\ast) \equiv -v(\mu,z_\ast) \Phi(\mu,z_\ast).
\label{krat4}
\end{equation}
Since the multiplicity of $\mu=\mu_\ast$ for $\Phi_\mu'(\mu,z_\ast)$ equals $k-1$ it follows from (\ref{krat4}) that its left-hand side is divisible by $(\mu-\mu_\ast)^k$ while one of its factors is divisible at most by $(\mu-\mu_\ast)^{k-1}$. Consequently, $u(\mu,z_\ast)$ is divisible by $\mu-\mu_\ast$ and hence $u(\mu_\ast,z_\ast)=0$.
Therefore, in any case, the condition (\ref{22*}) implies that $u(\mu_\ast,z_\ast)=0$. Formula (\ref{krat2}) yields then that ${\cal F}'(z_\ast)=v(\mu_\ast,z_\ast)\displaystyle{\left. \partial \Phi /\partial z \right|_{(\mu_\ast,z_\ast)}}$ and provided that $z_\ast$ is a simple zero for ${\cal F}(z)$, one has ${\cal F}'(z_\ast) \ne 0$ which results in $\displaystyle{\left. \partial \Phi / \partial z \right|_{(\mu_\ast,z_\ast)}} \ne 0$. To obtain the expression for the last derivative, let us differentiate the determinantal representation (\ref{fi})
\begin{eqnarray*}
\frac{\partial \Phi}{\partial z} &=& \mu^k \left|
\begin{array}{ccc}
{\bf A} & B & {\bf C}\\
\mathbb{O} & -\mu & \mathbb{O}\\
{\bf C}^T & \mathbb{O} & \displaystyle{-\frac{1}{\mu} }{\bf G}
\end{array} \right| = - \mu^{k+1} \left|
\begin{array}{cc}
{\bf A} & {\bf C}\\
{\bf C}^T & \displaystyle{-\frac{1}{\mu}} {\bf G}
\end{array} \right| \\
&=&- \mu^{k+1} \det ({\bf A}) \det \left( -\frac{1}{\mu} {\bf G} - {\bf C}^T{\bf A}^{-1}{\bf C} \right)\\
&=& (-1)^{k+1} \mu \det ({\bf A}) \det ({\bf G} + \mu {\bf C}^T{\bf A}^{-1}{\bf C})=(-1)^{k+1} \mu \det ({\bf A}) \det ({\bf M}).
\end{eqnarray*}
Since $\partial \Phi / \partial z \ne 0$ for $\mu=\mu_\ast$, $z=z_\ast$, the matrix ${\bf M}$ should be nonsingular for these values. \qed

\begin{corollary} If the system of columns $C_1, \dots, C_k$ is orthonormal then, by transforming the determinant in (\ref{eq8}), one can diminish its order: the expression under discriminant can be reduced into
\begin{equation}
\left|
\begin{array}{cc}
{\bf A}_1-\mu {\bf C}{\bf C}^T & B_1 \\
B_1^T & -1+\mu z
\end{array}
\right|.
\label{eq9}
\end{equation}
\end{corollary}

\begin{corollary}
Let $H \in \mathbb R^k$ be the given column. The condition
$$0 \le \left|
\begin{array}{ccc}
{\bf A}_1 & B_1 & {\bf C} \\
B_1^T & -1 & -H^T \\
{\bf C}^T & -H & \mathbb{O}
\end{array} \right| \times \left\{
\begin{array}{l}
(-1)^{k-1}, \mbox{ if } {\bf A}_1 \mbox{ is positive definite}, \\
(-1)^n, \mbox{ if } {\bf A}_1 \mbox{ is negative definite}
\end{array} \right.$$
is the necessary and sufficient one for the ellipsoid (\ref{eq1}) to intersect the affine subspace ${\bf C}^TX=H$. If this condition is not fulfilled then the square of the distance between the ellipsoid and the linear manifold equals the minimal positive zero of the polynomial
\begin{equation}
\mathcal F(z)=\mathcal D_\mu \left( \mu^k \left|
\begin{array}{ccc}
{\bf A}_1 & B_1 & {\bf C} \\
B_1^T & -1+\mu z & -H^T \\
{\bf C}^T & -H & \displaystyle{ \frac{1}{\mu} } {\bf G}
\end{array} \right| \right)
\label{lin_h}
\end{equation}
provided that this zero is not multiple one.
\end{corollary}
{\bf Proof} is similar to that of Theorem \ref{t1}. \qed

{\it Example.} Find the distance to the $x_1$-axis from the ellipsoid
$$7\,x_1^2+6\,x_2^2+5\,x_3^2-4\,x_1x_2-4\,x_2x_3-37\,x_1-12\,x_2+3\,x_3+54=0 \enspace .$$

{\it Solution.} One may choose here $C_1=[0,1,0]^T, C_2=[0,0,1]^T$, then the determinant (\ref{eq9}) takes the form
$$\left |
\begin{array}{cccc}
-7/54 & 1/27 & 0 & 37/108 \\
1/27 & -1/9-\mu & 1/27 & 1/9 \\
0 & 1/27 & -5/54-\mu & -1/36 \\
37/108 & 1/9 & -1/36 & -1+\mu z
\end{array}
\right |.$$
Equation (\ref{eq8})
\begin{eqnarray*}
{\cal F}(z) &=& 516019098077413632 \, z^4- 15034745857812486912 \, z^3  \\
&+& \,95300876926947983328 \, z^2 - 421036780846089455856 \, z  \\
&+& \, 237447832908365535785=0
\end{eqnarray*}
has two real zeros: $z_1 \approx 0.05712805$ and $z_2 \approx 22.54560673$. Hence, the distance equals $\sqrt{z_1}\approx 0.23901475$.

\begin{corollary} The square of the distance from the point $X_0$ to the ellipsoid (\ref{eq1}) coincides with the minimal positive zero of the equation
\begin{equation}
{\cal F}(z) \stackrel{def}{=} {\cal D}_\mu \left( \det \left( \left[
\begin{array}{cc}
{\bf A}_1 & B_1 \\
B_1^T & -1
\end{array} \right] +\mu \left[
\begin{array}{cc}
-{\bf I} & X_0 \\
X_0^T & z-X_0^TX_0
\end{array} \right] \right) \right) =0 \label{dispoint}
\end{equation}
provided that this zero is not a multiple one and $X_0^T{\bf A}_1X_0 +2B_1^TX_0 -1 \ne 0.$

The square of the distance from the origin $X= {\mathbb O}$ to the ellipsoid (\ref{eq1}) coincides with the minimal positive zero of the equation
\begin{equation}
{\cal F}(z) \stackrel{def}{=} {\cal D}_\mu \left(
(\mu z-1) \det ({\bf A}_1-\mu {\bf I})-B_1^T \adj ({\bf A}_1 - \mu {\bf I}) B_1 \right)=0
\label{eq10}
\end{equation}
provided that this zero is not a multiple one. Here $\adj$ stands for the adjoint matrix.
\end{corollary}

{\it Remark.} For large $n$, one can compute $\det ({\bf A}_1 -\mu {\bf I})$ and $\adj ({\bf A}_1 - \mu {\bf I})$ simultaneously with the aid of the Leverrier-Faddeev method \cite{b7}.

{\it Remark.}  For the case $B_1={\mathbb O}$, one gets ${\cal F}(z)\equiv {\cal D}(f)\left[ z^nf(1/z) \right]^2$ with $f(\mu)=\det ({\bf A}_1 -\mu {\bf I})$. This corres\-ponds to the well-known result that the distance to the ellipsoid $X^T{\bf A}_1X=1$ from its center coincides with the square root of the reciprocal of the largest eigenvalue of the matrix ${\bf A}_1$.

We exploit the result of the last corollary to elucidate the importance of the simplicity restriction imposed on the minimal positive zero for ${\cal F}(z)$; this assumption will also appear in the foregoing results.

{\it Example.} Find the polynomial (\ref{dispoint}) for the ellipse $x^2/4+y^2=1$ and the point $(x_0, y_0)$.

{\it Solution.} The polynomial ${\cal F}(z)$ from (\ref{dispoint}) for the ellipse $x^2/a^2+y^2/b^2=1$ is computed as
\begin{eqnarray*}
{\mathcal F}(z)&=&{\mathcal D}_{\mu} \left( \mu^3-\left\{a^2+b^2-x_0^2-y_0^2+z \right\}\mu^2 \phantom{\frac{1}{x}}\right.\\
&+&\left. \left\{-a^2b^2\left(\frac{x_0^2}{a^2}+\frac{y_0^2}{b^2} -1 \right)+z(a^2+b^2) \right\}\mu -  a^2b^2z \right),
\end{eqnarray*}
which for our particular case $a=2, \, b=1$ yields (up to a factor $1/256$)
\begin{eqnarray}
{\mathcal F}(z)&=&9\,z^4-6(2\,x_0^2+7\,y_0^2+15)\,z^3 \nonumber \\
&+& (-2\,x_0^4+73\,y_0^4+62\,x_0^2y_0^2-90\,x_0^2+270\,y_0^2+297)\,z^2 \nonumber \\
&+& (-56\,y_0^6-360\,y_0^2-62\,x_0^4-248\,y_0^4+4\,x_0^6+270\,x_0^2 \nonumber \\
&-& 90\,x_0^2y_0^4-30\,x_0^4y_0^2+140\,x_0^2y_0^2-360)\,z \nonumber \\
&+& 4(x_0^4+2\,x_0^2y_0^2+y_0^4-6\,x_0^2+6\,y_0^2+9)(x_0^2/4+y_0^2-1)^2 \ . \label{Fex}
\end{eqnarray}

Let us evaluate its zeros for $y_0=0$, i.e. for the points in $x-$axis:
$${\mathcal F}(z)= (z-(x_0-2)^2)(z-(x_0+2)^2)(3\,z - (3-x_0^2))^2.$$
Multiple zero $z_2=1-x_0^2/3$ is positive for $x_0 \in [0, \sqrt{3} \,[$. Moreover, for these values of $x_0$, zero $z_2$ is the minimal one for ${\cal F}(z)$. Nevertheless, for $x_0>3/2$, the square of the distance from $(x_0,0)$ to the ellipse equals $z_1=(x_0-2)^2.$ Explanation of this phenomenon is as follows: the multiple zero $z_2$ corresponds to the pair of points $(4 \, x_0/3, \, \pm \sqrt{1-4\, x_0^2/9})$ in the ellipse. These points are real for $x_0 \in [0, 3/2 \,[$ and imaginary (complex-conjugate) for $x_0>3/2$.

To conclude this example, let us illuminate the relationship of the stated metric problem to an ancient one concerning conic sections. Let us estimate the number of real zeros of the polynomial (\ref{Fex}). For this purpose, the sign of discriminant of polynomial is significant. One has
$$\Psi(x_0,y_0)= \mathcal D_z ( \mathcal F(z)) \equiv -\frac{9\, x_0^2y_0^2}{274877906944}((4\,x_0^2+y_0^2-9)^3 +972\,x_0^2y_0^2)^3.$$
\begin{figure}
\includegraphics{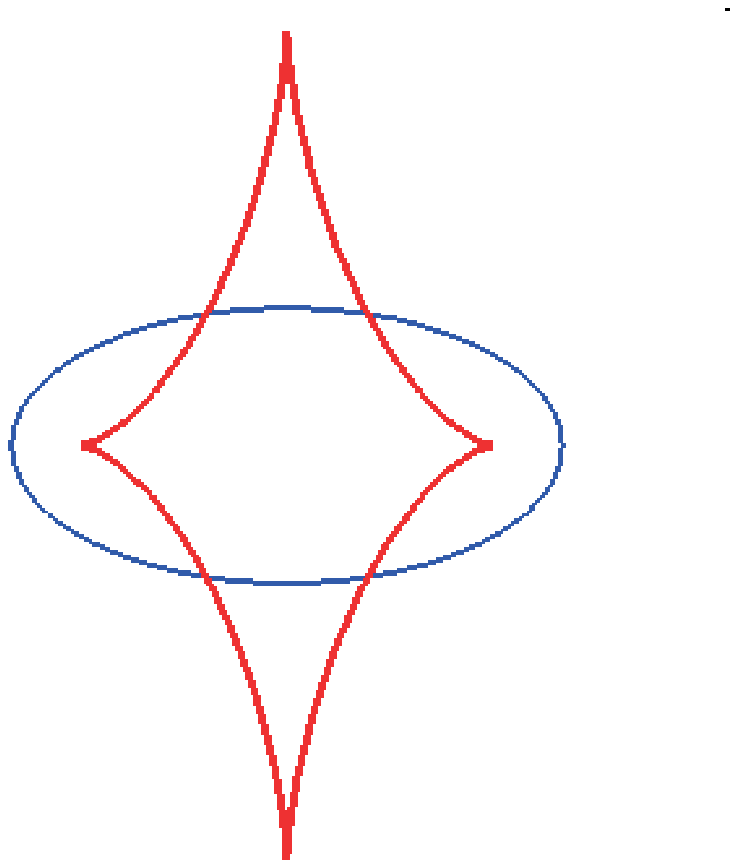}
\caption{}
\label{figa}
\end{figure}
Drawn in the $(x_0,y_0)-$plane, the curve $\Psi(x_0,y_0)=0$ consists of three branches: the coordinate axes and the curve known as astroid (marked in red in Fig. \ref{figa}). The latter was first treated by Apollonius in the 3rd century BC, in connection with the problem of finding the number of normals drawn from the given point to the ellipse. In terms of the zeros of polynomial (\ref{Fex}), the solution is as follows: for the points $(x_0,y_0)$ inside the astroid the polynomial ${\cal F}(z)$ posseses four real zeros, for those outside -- two. The exceptional points lie in the axes: one gets four real zeros for corresponding ${\cal F}(z)$ (with two of them becoming negative outside astroid).

To complete the present section, we provide estimations for degrees of polynomials ${\cal F}(z)$ appeared in the above results.

\begin{theorem} For the polynomial from (\ref{eq8}), one has generically $\deg {\cal F} =2k.$
\end{theorem}

{\bf Proof.} For simplicity, we will treat the case where the columns $C_1, \dots, C_k$ are orthonormal. We expand first the polynomial under the discriminant sign in powers of $z$:
$$\Phi(\mu,z)=z \mu^{k+1} \left|
\begin{array}{cc}
{\bf A}_1 & {\bf C}\\
{\bf C}^T & 1/\mu \, {\bf I}
\end{array} \right| + \mu^k \left|
\begin{array}{ccc}
{\bf A}_1 & B_1 & {\bf C} \\
B_1^T & -1 & \mathbb{O} \\
{\bf C}^T & \mathbb{O} & 1/\mu \, {\bf I}
\end{array} \right|.$$
Here ${\bf I}$ stands for the identity matrix of order $k$. The leading term of ${\cal F}(z)={\cal D}_\mu (\Phi(\mu,z))$ coincides with
$${\cal D}_\mu \left( z \mu^{k+1} \left|
\begin{array}{cc}
{\bf A}_1 & {\bf C} \\
{\bf C}^T & 1/\mu \, {\bf I}
\end{array} \right| \right).$$
In order to evaluate the degree of the last expression w.r.t. variable $z$, we may exploit the formula (\ref{unidiskcor1}). For this aim, it is necessary to find the degree of the polynomial under the discriminant sign w.r.t. $\mu$. Application of Schur formula (\ref{Schur}) in the way corresponding to (\ref{eq9}) yields
$$z \mu^{k+1} \left|
\begin{array}{cc}
{\bf A}_1 & {\bf C}\\
{\bf C}^T & 1/\mu \, {\bf I}
\end{array} \right| \equiv z \mu \det({\bf A}_1 - \mu {\bf C} {\bf C}^T)$$
which is not useful for our purpose since the matrix ${\bf C}{\bf C}^T$ is singular if $k < n$. Let us use Schur formula in an alternative way:
$$\equiv z \mu \det {\bf A}_1 \det ({\bf I} - \mu {\bf C}^T {\bf A}_1^{-1} {\bf C}).$$
The last determinant is of the order $k$ with all of its entries depending linearly on $\mu$. We expand it in decreasing powers of $\mu$:
$$\equiv (-1)^{k+1} z \mu^{k+1} \det {\bf A}_1 \det ({\bf C}^T{\bf A}_1^{-1} {\bf C}) + \dots$$
Since, by the assumption, matrix ${\bf A}_1$ from the equation (\ref{eq1}) provides an ellipsoid, all the matrices ${\bf A}_1$, ${\bf A}_1^{-1}$ and ${\bf C}^T {\bf A}_1^{-1} {\bf C}$ are sign-definite. Therefore, their determinants do not vanish and the leading term of ${\cal F}(z)$ equals generically
$$z^{2k}(\det {\bf A}_1)^{2k} {\cal D}_\mu (\det ({\bf I} - \mu {\bf C}^T{\bf A}_1^{-1} {\bf C})) \equiv z^{2k}(\det {\bf A}_1)^2 {\cal D}_\mu(\det ({\bf A}_1 -\mu {\bf C}{\bf C}^T)).$$ \qed

\begin{corollary} For the polynomial from (\ref{dispoint}), the leading term equals generically to
$$z^{2n} (\det {\bf A}_1)^2 {\cal D}_\mu (\det ({\bf A}_1 - \mu {\bf I})).$$
\end{corollary}

\section{Distance between quadrics}
\label{DisQQ}
Consider first the case of surfaces (\ref{eq1}) and (\ref{eq3}) centered at the origin: $B_1= {\mathbb O}, \ B_2={\mathbb O}$.

\begin{theorem} The surfaces $X^T{\bf A}_1X=1$ and $X^T{\bf A}_2X=1$ intersect iff the matrix ${\bf A}_1 - {\bf A}_2$ is not sign-definite. If this condition is not satisfied then the value $d^2$ coincides with the minimal positive zero of the equation
\begin{equation}
{\cal F}(z) \stackrel{def}{=} {\cal D}_\lambda (\det (\lambda {\bf A}_1 + (z- \lambda) {\bf A}_2 - \lambda (z-\lambda) {\bf A}_1 {\bf A}_2))=0
\label{eq11}
\end{equation}
provided that this zero is not a multiple one.
\label{t2}
\end{theorem}

{\bf Proof.}  {\bf I}. The intersection condition can be found as an exercise in the problem book \cite{b8}. We just repeat here the arguments.

Since the equation $X^T{\bf A}_1X=1$ provides an ellipsoid, the matrix ${\bf A}_1$ is positively definite.

Let there exist a point $X=X_0 \in \mathbb{R}^n$ such that $X_0^T {\bf A}_1X_0=1$ and $X_0^T {\bf A}_2X_0=1$. Thus $X_0^T ( {\bf A}_1 -{\bf A}_2) X_0=0$ for $X_0 \ne \mathbb{O}$. Therefore, $ {\bf A}_1 - {\bf A}_2$ is not sign-definite.

Conversely, if $ {\bf A}_1 - {\bf A}_2$ is not sign-definite then there exists $X_0 \ne \mathbb{O}$ such that $X_0^T ( {\bf A}_1 -{\bf A}_2) X_0=0$ or, alternatively, $X_0^T{\bf A}_1X_0=X_0^T{\bf A}_2X_0$. Multiply the latter by a scalar $t^2$ with $t \in \mathbb{R}$: $t^2X_0^T{\bf A}_1X_0=t^2X_0^T{\bf A}_2X_0$. Set
$t=1/\sqrt{X_0^T{\bf A}_1X_0}$ (the radicand is positive due to the positive definiteness of ${\bf A}_1$). The point $X=tX_0$ is an intersection point of both manifolds since
$$X^T{\bf A}_1X=t^2X_0^T{\bf A}_1X_0=1 \mbox{ and } X^T{\bf A}_2X=t^2X_0^T{\bf A}_2X_0=t^2X_0^T{\bf A}_1X_0=1.$$

{\bf II}. If the intersection condition is not valid, then the distance problem becomes nontrivial and we apply the Lagrange multipliers method for the objective function in the form
$$(X-Y)^T(X-Y)-\lambda_1(X^T{\bf A}_1X-1)-\lambda_2(Y^T{\bf A}_2Y-1).$$
The corresponding system of algebraic equations is as follows
\begin{eqnarray}
&&X-Y-\lambda_1{\bf A}_1X=\mathbb{O}, \, -X+Y-\lambda_2{\bf A}_2Y=\mathbb{O}, \label{z1}\\
&&X^T{\bf A}_1X=1, \, Y^T{\bf A}_2Y=1. \label{z2}
\end{eqnarray}
This system yields
\begin{eqnarray}
&& ( \lambda_1 \lambda_2 {\bf A}_2 {\bf A}_1 - \lambda_1 {\bf A}_1 - \lambda_2 {\bf A}_2) X ={\mathbb O}, \label{43} \\
&& (\lambda_1 \lambda_2 {\bf A}_1 {\bf A}_2 -\lambda_1 {\bf A}_1- \lambda_2 {\bf A}_2) Y = {\mathbb O}, \label{44}
\end{eqnarray}
and
\begin{eqnarray}
&&\lambda_1 {\bf A}_1X+\lambda_2 {\bf A}_2Y=\mathbb O, \label{47}\\
&&X-Y=\lambda_1 {\bf A}_1X. \label{48}
\end{eqnarray}
Matrices of the systems (\ref{43}) and (\ref{44}) differs only by transposition, and therefore the determinants of these matrices are equal. Their common value should be just $0$ due to the fact that we are looking for nontrivial solutions of homogeneons systems:
\begin{equation}
\det(\lambda_1 \lambda_2 {\bf A}_1 {\bf A}_2 -\lambda_1 {\bf A}_1 - \lambda_2 {\bf A}_2)=0. \label{49}
\end{equation}
Let us introduce the matrix
\begin{equation}
{\bf M} \stackrel{def}{=} {\bf I}-\displaystyle{\frac{1}{\lambda_1}} {\bf A}_1^{-1} - \displaystyle{\frac{1}{\lambda_2}} {\bf A}_2^{-1}
\label{z3}
\end{equation}
and the vector
\begin{equation}
Z \stackrel{def}{=} X-Y. \label{53}
\end{equation}
Using this notation, the equations (\ref{43}) and (\ref{44}) can be rewritten into equivalent form
\begin{equation}
{\bf M}Z= \mathbb O \quad \Leftrightarrow \quad Z=\left( \frac{1}{\lambda_1} {\bf A}_1^{-1} + \frac{1}{\lambda_2} {\bf A}_2^{-1} \right) Z, \label{54}
\end{equation}
while the conditions (\ref{z2}) in the form
\begin{equation}
\frac{1}{\lambda_j^2}Z^T{\bf A}_jZ=1 \mbox{ for } j \in \{1,2\}. \label{55}
\end{equation}
Let us introduce a new variable $z$ responsible for the critical values of the distance function
\begin{eqnarray}
z&=&(X-Y)^T(X-Y) \stackrel{(\ref{53})}{=}Z^TZ   \label{56}\\
&\stackrel{(\ref{54})}{=}&\frac{1}{\lambda_1}Z^T{\bf A}_1^{-1}Z+\frac{1}{\lambda_2}Z^T{\bf A}_2^{-1}Z \stackrel{(\ref{55})}{=} \lambda_1 + \lambda_2. \nonumber
\end{eqnarray}
Thus, we have eliminated the variables $X$ and $Y$ from the system (\ref{z1})-(\ref{z2}) with the resulting equations assuming the form (\ref{49}) and (\ref{56}). To deduce an extra equation, one should start with the identity
$${\bf M} \cdot \adj ( {\bf M}) = {\bf I} \cdot \det {\bf M}.$$
By differentiation this as to $\lambda_j$, one obtains
$$\frac{\partial {\bf M}}{\partial \lambda_j}\, \adj ( {{\bf M}}) + {\bf M} \,\frac{\partial \adj ( {{\bf M}})}{\partial \lambda_j} \equiv \frac{\partial \det {\bf M} }{\partial \lambda_j} \, {\bf I}.$$
Multiply this by $Z^T$ from the left-hand side and by $Z$ from the right-hand side, with $Z$ standing for any nontrivial solution to the system (\ref{54}):
\begin{equation}
Z^T\,\frac{\partial {\bf M}}{\partial \lambda_j}\, \adj ( {{\bf M}}) Z+Z^T{\bf M} \,\frac{\partial \adj ( {{\bf M}})}{\partial \lambda_j}\,Z \equiv \frac{\partial \det {\bf M}}{\partial \lambda_j} \,Z^TZ. \label{57}
\end{equation}
Taking into account (\ref{54}) and symmetry of the matrix ${\bf M}$, one arrives at
$$Z^T{\bf M}=({\bf M}Z)^T=\mathbb{O},$$
and therefore identity (\ref{57}) turns to
\begin{equation}
Z^T \frac{\partial {\bf M}}{\partial \lambda_j} \adj ( {\bf M}) Z=\frac{\partial \det {\bf M}}{\partial \lambda_j} Z^TZ, \label{z3k}
\end{equation}
or, in view of (\ref{z3}):
\begin{equation}
\frac{1}{\lambda_j^2}Z^T{\bf A}_j^{-1}\adj ( {\bf M}) Z=\frac{\partial \det {\bf M}}{\partial \lambda_j} Z^TZ. \label{z3n}
\end{equation}
Now, our aim is to prove that
\begin{equation}
\adj ( {\bf M}) Z =\gamma Z \label{58}
\end{equation}
for a certain scalar $\gamma$. Indeed,
$$\adj ( {\bf M}) {\bf M}Z= \mathbb{O} \quad \Leftrightarrow \quad {\bf M} (\adj ( {\bf M}) Z)=\mathbb{O}.$$
If $\rank ({\bf M}) =n-1$ then any solution $U$ to the system of homogeneons equations ${\bf M}U= \mathbb{O}$ should be equal just a multiple of $Z$; therefore
$$\adj ( {\bf M}) Z =\gamma Z.$$
The case $\rank ({\bf M}) <n-1$ is trivial since $\adj ( {\bf M}) = \mathbb{O}_{n \times n}$. (It can be proved that in any case
$\gamma= {\bf M}_{11}+{\bf M}_{22}+ \dots +{\bf M}_{nn}$ with ${\bf M}_{jj}$ standing for the cofactor to the corresponding entry of ${\bf M}$.)

Hence, the formula (\ref{z3n}) is transformed into
$$\frac{\gamma}{\lambda_j^2}Z^T{\bf A}_j^{-1}Z=\frac{\partial \det {\bf M}}{\partial \lambda_j} \, Z^TZ,$$
wherefrom one can deduce (with the aid of (\ref{55})) that
\begin{equation}
\frac{\partial \det {\bf M}}{\partial \lambda_1} = \frac{\partial \det {\bf M}}{\partial \lambda_2}. \label{59}
\end{equation}

Recalling now that $\lambda_1$ and $\lambda_2$ are connected via condition (\ref{56}), we substitute $\lambda_1=z-\lambda_2$ into (\ref{59}) and obtain
$$\frac{\partial \det {\bf M}}{\partial \lambda_2} \, \frac{d \lambda_2}{d \lambda_1} = \frac{\partial \det {\bf M}}{\partial \lambda_2} \quad \Rightarrow \quad \frac{\partial \det {\bf M}}{\partial \lambda_2}=0.$$
Thus, the process of elimination of variables from the system (\ref{z2})--(\ref{44}) and (\ref{56}) terminates when we get the two equations: the first one is
\begin{equation}
\det \left({\bf I} - \frac{1}{z-\lambda_2} {\bf A}_1^{-1} - \frac{1}{\lambda_2} {\bf A}_2^{-1} \right) =0, \label{60}
\end{equation}
while the second is obtained from this by differentiation its left-hand side as to $\lambda_2$. Elimination of $\lambda_2$ from these equations can be performed in the traditional manner, i.e. via discriminant. Utilizing the result of Corollary \ref{colfrac} from Sect. \ref{AlgPr}, we turn from the rational functions to polynomial ones. Multiplication of (\ref{60}) by $\det({\bf A}_1 {\bf A}_2)$ and substitution $\lambda=1/\lambda_2$ completes the proof.

{\bf III}. To find the nearest points on the quadrics we suggest the following approach. Once the real zero $z=z_\ast$ of (\ref{eq11}) is evaluated, one can find the corresponding value $\lambda=\lambda_\ast$ which is a multiple zero for
$$G(\lambda,z_\ast) \stackrel{def}{=} \det (\lambda {\bf A}_1 +(z_\ast- \lambda){\bf A}_2 -\lambda (z_\ast-\lambda) {\bf A}_2{\bf A}_1).$$
Under the assumption of the theorem, this zero is unique and can be expressed rationally in terms of the coefficients of the polynomial $G(\lambda, z_\ast)$ with the aid of (\ref{eq7}). Futhermore, the coordinate column $X_\ast$ of the point on the quadric $X^T{\bf A}_1X=1$ is a solution for the system of homogeneons equations
\begin{equation}
(\lambda_\ast {\bf A}_1 +(z_\ast- \lambda_\ast){\bf A}_2-\lambda_\ast (z_\ast-\lambda_\ast) {\bf A}_2{\bf A}_1)X=\mathbb{O}, \label{61}
\end{equation}
which possesses an infinite number of solutions since its determinant vanishes. From the solution set one should choose a representative satisfying the condition $X^T{\bf A}_1X=1$. Due to symmetry of the problem, there exists a pair of such solutions.

Similarly, the coordinate column for the point in the second quadric satisfies the system
\begin{equation}
(\lambda_{\ast} {\bf A}_1 +(z_{\ast}-\lambda_{\ast}){\bf A}_2 - \lambda_{\ast} (z_{\ast}-\lambda_{\ast}) {\bf A}_1{\bf A}_2) Y   = \mathbb O . \label{62}
\end{equation}

Recall that the matrices of the system (\ref{61}) and (\ref{62}) differ only by transposition and in order to solve both systems (\ref{61}) and (\ref{62}) it suffice to treat the rows and the columns of the matrix adjoint to the matrix
$${\bf M}_\ast= \lambda_{\ast} {\bf A}_1 +(z_{\ast}-\lambda_{\ast}){\bf A}_2 - \lambda_{\ast} (z_{\ast}-\lambda_{\ast}) {\bf A}_2{\bf A}_1 .$$
Indeed, $X_\ast^T$ equals just a multiple of any nonzero row of the matrix $\adj( {\bf M}_\ast)$ while $Y_\ast$ coincides with a multiple of any nonzero column of $\adj( {\bf M}_\ast)$. By a suitable selection of the mentioned multipliers, one can provide the fulfilment of the conditions $X^T{\bf A}_1X=1$ and $Y^T{\bf A}_2Y=1$. The obtained pairs of points should be adjusted according to the condition
$$(X_\ast-Y_\ast)^T(X_\ast-Y_\ast)=z_{\ast}.$$ \qed

{\it Example.} Find the distance between the ellipses
$$10\,x_1^2-12\,x_1x_2+8\,x_2^2=1  \mbox{ and } x_1^2+x_1x_2+x_2^2=1 \ . $$

{\it Solution.} Here
$${\bf A}_1= \left(
\begin{array}{cc}
10 & - 6 \\
-6 & 8
\end{array}
\right), \quad
{\bf A}_2=
\left(
\begin{array}{cc}
1 & \frac{1}{2} \\
\frac{1}{2}  & 1
\end{array}
\right)$$
and the matrix ${\bf A}_1-{\bf A}_2$ is positive definite. Thus the ellipses do not intersect.

Compose the determinant from Theorem \ref{t2}.
\begin{eqnarray}
G(\lambda,z)&=&\det(\lambda {\bf A}_1 + (z- \lambda) {\bf A}_2 - \lambda (z-\lambda) {\bf A}_1 {\bf A}_2) \nonumber \\
&=& \left|
\begin{array}{cc}
7\,\lambda^2+(-7z+9)\lambda+z  & -2\lambda^2+\left(2\,z-\displaystyle{\frac{13}{2}}\right)\lambda+\displaystyle{\frac{1}{2}\,z}  \\
& \\
-\lambda^2+\displaystyle{\left(z-\frac{13}{2}\right)\lambda+\frac{1}{2}\,z }& 5\lambda^2+(-5z+7)\lambda+z
\end{array}
\right| \label{mM} \\
&=& 33\,\lambda^4+\left(-66z+\frac{149}{2}\right)\lambda^3+\left(33\,z^2-61\,z+\frac{83}{4}\right)\lambda^2 \nonumber \\
&+&\left(-\frac{27}{2}z^2+\frac{45}{2}z\right)\lambda+\frac{3}{4}\,z^2 \nonumber
\end{eqnarray}
The discriminant of this polynomial w.r.t. $ \lambda $ equals
$${\mathcal F}(z)=\frac{3}{16}\,z^2 ( 936086976 \, z^6- 10969697376 \,z^5+ 50706209664 \, z^4 $$
$$ - 115515184664 \, z^3+ 130176444432 \, z^2 - 59826725574 \,z+ 2866271785 ) \ .$$
Its positive zeros are as follows:
$$z_\ast \approx 0.053945666,\ 1.3340583883,\ 1.95921364,\ 2.8785867381 \ .$$
Hence, $d=\sqrt{z_\ast} \approx 0.23226206 $.

To find the nearest points on the given ellipses, establish first the multiple zero of the polynomial $G(\lambda, z_\ast)$ with the aid of Theorem \ref{t21}:
$$\lambda={\displaystyle -\frac{-{\displaystyle 725274}\,z^5+{\displaystyle 1455894}\,z^4+{\displaystyle \frac{11286981}{2}} z^3- {\displaystyle \frac{26486523}{2}} z^2+{\displaystyle \frac{42000075}{8}}z}
{{\displaystyle 17591706}\,z^4-{\displaystyle 109992894}\,z^3+{\displaystyle \frac{450450691}{2}}z^2- {\displaystyle \frac{315606253}{2}}z+ {\displaystyle \frac{77466805}{8}}}}.$$
Substitution $z=z_\ast$ yields $\lambda=\lambda_\ast \approx -0.13576051$.

Secondly, for the obtained pair of values $z_\ast$ and $\lambda_\ast$ the determinant (\ref{mM}) vanishes and therefore both systems (\ref{61}) and (\ref{62}):
$${\bf M}_\ast X=\mathbb O, \ {\bf M}_\ast^T Y=\mathbb O$$
posses nontrivial solutions. To find these solutions, take the first row and the first column of the matrix $\adj( {\bf M}_\ast)$

$$X=\left[ \left.
\begin{array}{c}
2\lambda^2-(2\,z-\frac{13}{2})\lambda-\frac{1}{2}z  \\
\\
7\,\lambda^2+(-7z+9)\lambda+z
\end{array}
\right] \right|_{z=z_{\ast}, \lambda= \lambda_{\ast}}
\approx
\left(
\begin{array}{c}
-0.8579069 \\
\\
-0.9876166
\end{array}
\right),$$
$$Y=\left[ \left.
\begin{array}{c}
\lambda^2-(z-\frac{13}{2})\lambda-\frac{1}{2}z  \\ [2ex]
7\,\lambda^2+(-7z+9)\lambda+z
\end{array}
\right] \right|_{z=z_{\ast}, \lambda= \lambda_{\ast}}
\approx
\left(
\begin{array}{c}
-0.8836615 \\ [2ex]
-0.9876166
\end{array}
\right).$$
Each such point defines a line passing through the origin. To find intersection points with the corresponding ellipses, one should make normalization
$$X_{\ast}=\frac{\pm X}{\sqrt{X^T{\bf A}_1 X}} \approx \pm
\left(
\begin{array}{c}
-0.3838312 \\
-0.4418639
\end{array}
\right), \quad
Y_{\ast} =\frac{\pm Y}{\sqrt{Y^T{\bf A}_2 Y}} \approx \pm
\left(
\begin{array}{c}
-0.5449964 \\
-0.6091105
\end{array}
\right).$$
These formulas provide two pairs of nearest points in the ellipses.

Let us now treat the general case of manifolds position.
\begin{theorem} The surfaces $X^T{\bf A}_1X+2B_1^TX-1=0$ and $X^T{\bf A}_2X+2B_2^TX-1=0$ intersect iff among the real zeros of the equation
$$\Phi (z) \stackrel{def}{=} {\cal D}_\lambda \left( \det \left( \left[
\begin{array}{cc}
{\bf A}_2 & B_2\\
B_2^T & -1-z
\end{array} \right] - \lambda \left[
\begin{array}{cc}
{\bf A}_1 & B_1\\
B_1^T & -1
\end{array} \right] \right) \right) =0$$
there are the values of different signs or $0$. If this condition is not fulfilled then the value $d^2$ coincides with the minimal positive zero of the equation
\begin{eqnarray}
{\cal F}(z) &\stackrel{def}{=}&
\label{eq12}
{\cal D}_{\mu_1, \mu_2} \left( \det \left( \mu_1 \left[
\begin{array}{cc}
{\bf A}_1 & B_1\\
B_1^T & -1
\end{array} \right]  +\,\mu_2 \left[
\begin{array}{cc}
{\bf A}_2 & B_2\\
B_2^T & -1
\end{array} \right] \right. \right.  \\
&& - \left. \left.  \left[
\begin{array}{cc}
{\bf A}_2 {\bf A}_1 & {\bf A}_2 B_1\\
B_2^T {\bf A}_1 & B_2^TB_1 - \mu_1 \mu_2 z
\end{array} \right] \right) \right) =0 \nonumber
\end{eqnarray}
provided that this zero is not a multiple one.
\label{t3}
\end{theorem}

{\bf Proof.} We sketch it as it is similar to that of Theorem \ref{t1}. Intersection condition is a result of the following considerations. Extrema of the function $X^T{\bf A}_2X+2B_2^TX-1$ on the ellipsoid (\ref{eq1}) are all of the similar sign iff the surfaces (\ref{eq1}) and (\ref{eq3}) do not intersect. We state the problem of
finding the extremal values of $V(X)=X^T{\bf A}_2X+2B_2^TX-1$ subject to (\ref{eq1}), then apply the Lagrange multipliers method and finally eliminate all the variables except for $z$ from the obtained algebraic system coupled with the equation $X^T{\bf A}_2X+2B_2^TX-1-z=0$.

To prove the second part of the theorem, we take the matrix ${\bf M}$ defined by (\ref{z3}), while
$$Q \stackrel{def}{=} -{\bf A}_1^{-1}B_1 + {\bf A}_2^{-1}B_2,$$
and transform the equations of the system (\ref{eq4}) into
\begin{eqnarray}
& &X=-{\bf A}_1^{-1}B_1+\displaystyle{\frac{1}{\lambda_1}{\bf A}_1^{-1}{\bf M}^{-1}Q, \quad
Y=-{\bf A}_2^{-1}B_2-\frac{1}{\lambda_2}{\bf A}_2^{-1}{\bf M}^{-1}Q},
\label{eq13} \\
& &-B_j^T{\bf A}_j^{-1}B_j+\displaystyle{\frac{1}{\lambda_j^2}} Q^T{\bf M}^{-1}{\bf A}_j^{-1}
{\bf M}^{-1}Q-1=0 \mbox{ for } j \in \{ 1,2 \},
\label{eq14}\\
& &z-Q^T{\bf M}^{-2}Q=0.
\label{eq15}
\end{eqnarray}
On multiplying equations (\ref{eq14}) by $\lambda_j$ and using (\ref{eq15}), we deduce that
\begin{equation}
-\lambda_1 B_1^T{\bf A}_1^{-1}B_1-\lambda_2 B_2^T{\bf A}_2^{-1}B_2- Q^T{\bf M}^{-1}Q -\lambda_1 - \lambda_2+z=0.
\label{eq16}
\end{equation}
It can be verified that the derivative of the left-hand side of (\ref{eq16}) with respect to $\lambda_j$ coincides with that one of (\ref{eq14}). Substitution
$\mu_1=1/\lambda_2, \ \mu_2=1/\lambda_1$ and the use of Schur formula (\ref{Schur}) enable one to reduce (\ref{eq16}) to the determinantal representation from (\ref{eq12}). \qed

{\it Example.} Find the distance between the ellipsoids
$$7\,x_1^2+6\,x_2^2+5\,x_3^2-4\,x_1x_2-4\,x_2x_3-37\,x_1-12\,x_2+3\,x_3+54=0$$
$$\mbox{and } 189\,x_1^2+x_2^2+189\,x_3^2+2\,x_1x_3-x_2x_3-27=0$$
and establish the coordinates of their nearest points.

{\it Solution.} Intersection condition from Theorem \ref{t3} is not satisfied: the 6th degree polynomial $\Phi(z)$ has all its real zeros positive. To compute the discriminant we use the result of Theorem \ref{tva} with the matrix $\mathfrak B$ of the order 16 constructed for the set (\ref{stepeni}). Its determinant is the 24th degree polynomial ${\cal F}(z)$ with integer coefficients of the orders up to $10^{188}$. It has eight positive zeros
$$z_1 \approx 1.3537785, \ z_2 \approx 3.5509348, \dots, z_8 \approx 111.7480312.$$ Thus, the distance between the given ellipsoids equals $\sqrt{z_1} \approx 1.1635198.$

For the obtained value of $z_1$, polynomial in $\mu_1$ and $\mu_2$ from (\ref{eq12}) possesses a multiple zero which can be expressed rationally in terms of $z_1$ with the aid of the minors of $\mathfrak B$ via formulas (\ref{eq7}). Substitution of the obtained values $\lambda_1 \approx 5.75593612, \ \lambda_2 \approx -0.45858332$ into (\ref{eq13}) yields the coordinates of the nearest points on the given ellipsoids:
$$X \approx [1.5203947, \ 1.5098600, \ 0.1262343]^T,$$ $$Y \approx [0.3610045, \ 1.4849072, \ 0.0315226]^T.$$

\section{Parameter dependent surfaces}
\label{DisPar}
The problem of distance estimation between moving objects in 3D space is of importance to astronomy, robotics and computer graphics. To illuminate the perspectives of the approach developed in the previous sections for such problems dealing with quadrics, we will treat the following problem.

Find the distance from the point $X_0 \in \mathbb R^n$ to the nearest point of the family of ellipsoids in $\mathbb R^n$
\begin{equation}
\left\{ X^T {\bf A}_1(t)X+2B_1^T(t)X-1=0 \, \bigg | \, t \in [a,b] \right\} \label{mov1}
\end{equation}
with the coefficients of ${\bf A}_1(t)$ and $B_1(t)$ polynomially dependent on the parameter $t$.
\begin{theorem} The square of the distance from $X_0$ to (\ref{mov1}) coincides with the minimal positive zero of one of the equations
$$\mathfrak F (z) \stackrel{def}{=} {\cal D}_t({\cal F}(z,t))=0, \ {\cal F}(z,a)=0, \ {\cal F}(z,b)=0.$$
Here ${\cal F}(z,t)$ is a polynomial (\ref{dispoint}) and the mentioned zero is not a multiple one.
\end{theorem}

In short: the stated problem can be solved with the aid of {\it iterated} discriminant.

{\bf Proof.} For any given value of $t$, the square of the distance from $X_0$ to the corresponding ellipsoid of the family (\ref{mov1}) is evaluated as a zero of the equation (\ref{dispoint})
\begin{equation}
{\cal F}(z,t)=0. \label{mov2}
\end{equation}
Due to imposed restrictions on the coefficients of the family, ${\cal F}$ is a polynomial function in $t$. Equation (\ref{mov2}) can be treated as defining an implicit function $z(t)$. It is known that zeros of a polynomial are continuously differentiable functions of the coefficients of this polynomial (except for the coefficient specializations annihilating the discriminant) \cite{b9}. Consequently, for any zero $z=z_\ast(t)$ of (\ref{mov2}) there exists the derivative $d z_\ast(t) / d t$. Differentiation of the equality ${\cal F}(z_\ast(t),t) \equiv 0$ with respect to $t$ results in
\begin{equation}
\frac{\partial \mathcal F}{\partial z} \cdot \frac{dz_\ast(t)}{dt} + \frac{\partial \mathcal F}{\partial t} \equiv 0, \label{mov3}
\end{equation}
here the partial derivatives are evaluated at $z=z_\ast(t)$.

For $t \in [a,b]$, the minimum of the function $z_\ast(t)$ is attained either at the end points of the interval or in the stationary point $t=\tilde{t}$ at which $dz_\ast / dt =0.$
In the latter case, it follows from (\ref{mov3}) that
\begin{equation}
\frac{\partial \mathcal F}{\partial t}=0 \label{mov4}
\end{equation}
at $t=\tilde{t}$. The two conditions (\ref{mov2}) and (\ref{mov4}) provide an algebraic system with respect to both variables $z$ and $t$. One can  eliminate the variable $t$ with the aid of discriminant. \qed

{\it Example.} Find the distance from the point $(-10, \, 10)$ to the family of ellipses
$$\left\{ \frac{(x-t)^2}{4}+\frac{(y-t(t-4))^2}{16}=1 \ \bigg | \ t \in \mathbb R \right\}.$$

{\it Solution.} We skip the expression for ${\cal F}(z,t)$.
\begin{eqnarray*}
{\cal D}_t({\cal F}(z,t))&=&z^4 (3\,z + 16888)^2 ( 9 \,z^2- 4080 \,z + 333376)^2 \\
&\times& ( 16777216 \,z^{12} - 24039653376 \,z^{11} + 15135396003840 \,z^{10} \\
&-& 5551772745220096 \,z^9 + 1322366761276505856 \,z^8 \\
&-& 215049198876048266976 \,z^7 + 24423380307243182292153 \,z^6 \\
&-& 1952292050779441220868024 \,z^5+ 109783307459960901970173936 \,z^4 \\
&-& 4304075084512715479517135104 \,z^3 \\
&+& 113714594973157300688449668864 \,z^2 \\
&-& 1830069428535779484150176987136 \,z \\
&+& 14265422520155306699255826485248)^3 \\
&\times& ( 2304 \,z^8 - 3774720 \,z^7 + 2645308000 \,z^6 - 1058624029488 \,z^5 \\
&+& 266900597798217 \,z^4 - 42785419475837458 \,z^3 \\
&+& 4100511694812810849 \,z^2 - 202905147887926860744 \,z \\
&+& 3648597980765724103824).
\end{eqnarray*}
The minimal positive zero of the last factor is $z_{\ast}\approx 37.70933565.$ The distance to the family equals $\sqrt{z_\ast} \approx 6.140792755$. One can find the ellipse of the family at which the distance is attained via the traditional application of the multiple zero evaluation formula (\ref{lam}): $t_{\ast} \approx -1.9680233599.$

\section{Conclusions}
\label{Conc}
We have treated the problem of distance evaluation between algebraic surfaces in ${\mathbb R}^n$ via inversion of the traditional approach:
\begin{center}
nearest points \ $\rightarrow$ \ distance \enspace .
\end{center}
This has been performed via  introduction of an extra variable responsible for the critical values of distance function and application of Elimination Theory methods. Such an approach was first suggested in \cite{b3} for the general polynomial optimization problem in $\mathbb R^n$. Its employment for the distance evaluation problem for quadrics has led to the result which happened to be surprisingly unpredictable for us: the discriminant is fully responsible for everything. With its help it is not only possible to deduce a univariate polynomial equation for the square of the distance but also to express (Theorem \ref{t3}) the necessary and sufficient condition for the intersection of the surfaces.

The major advantage of this approach over the traditional scheme is that the problem is reduced to evaluation of a single zero of a univariate algebraic equation instead of dealing with multidimensional constrained optimization problem. Moreover, introduction of an extra (distance) variable $z$ into the problem provides one with a nice (i.e. rational) parameterization of the nearest points coordinates.

Several problems have remained for further investigation, among them estimation of the degree of polynomial ${\cal F}(z)$ constructed for the problems of Sect. \ref{DisQQ}. The conjecture is that $\deg {\cal F}(z)=n(n+1)$ for ${\cal F}(z)$ from (\ref{eq11}) and that $\deg {\cal F}(z) =2n(n+1)$ for ${\cal F}(z)$ from (\ref{eq12}); with these estimations valid on excluding some extraneous factor (e.g. in case of (\ref{eq11}), this factor is just $z^{n(n-1)}$).

The proposed approach might be especially useful for the optimization problems connected with the parameter dependent surfaces like the one treated in Sect. \ref{DisPar} or for the multidimensional pattern recognition analysis.

\end{document}